\documentstyle[prd,aps,twocolumn,graphicx]{revtex}
\newcommand{\clp}{{\mathcal{P}}}
\newcommand{\cle}{{\mathcal{E}}}
\newcommand{\clb}{{\mathcal{B}}}
\newcommand{\TT}{\text{TT}}
\newcommand{\ud}{\text{d}}
\newcommand{\ue}{\text{e}}
\newcommand{\yg}{Y^{\mathrm{G}}}
\newcommand{\yc}{Y^{\mathrm{C}}}
\newcommand{\yp}{Y^{\mathrm{P}}}
\newcommand{\ag}{a^{\mathrm{G}}}
\newcommand{\ac}{a^{\mathrm{C}}}
\newcommand{\bg}{b^{\mathrm{G}}}
\newcommand{\bc}{b^{\mathrm{C}}}
\newcommand{\bp}{b^{\mathrm{P}}}
\newcommand{\blde}{\bbox{e}}
\newcommand{\bldm}{\bbox{m}}
\newcommand{\bldsig}{\bbox{\sigma}}
\newcommand{\sigco}{\bbox{\sigma}_{\text{co}}}
\newcommand{\sigcross}{\bbox{\sigma}_{\text{cross}}}
\newcommand{\etco}{\tilde{{\mathcal{E}}}_{\text{co}}}
\newcommand{\etcross}{\tilde{{\mathcal{E}}}_{\text{cross}}}
\begin{document}
\draft

\twocolumn[\hsize\textwidth\columnwidth\hsize\csname
@twocolumnfalse\endcsname

\title{All-Sky Convolution for Polarimetry Experiments}
\author{Anthony Challinor$^1$, Pablo Fosalba$^2$, Daniel Mortlock$^{1,3}$,
Mark Ashdown$^{1,3}$,\\
Benjamin Wandelt$^{4}$, and Krzysztof G\'{o}rski$^{5}$}
\address{${}^{1}$Astrophysics Group, Cavendish Laboratory, Madingley Road,
Cambridge CB3 0HE, UK\\
${}^{2}$Astrophysics Division, Space Science Dept. of ESA/ESTEC, NL-2200
AG Noordwijk, The Netherlands\\
${}^{3}$Institute of Astronomy, Madingley Road, Cambridge CB3 0HA, UK\\
${}^{4}$Department of Physics, Princeton University, Princeton, NJ 08540, USA\\
${}^{5}$European Southern Observatory, Garching bei M\"{u}nchen, Germany}
\date{\today}
\maketitle
\begin{abstract}
We discuss all-sky convolution of the instrument beam with the sky signal in
polarimetry experiments, such as the Planck mission
which will map the temperature anisotropy and polarization of the cosmic
microwave background (CMB). To account properly for stray light (from e.g.\ the
galaxy, sun, and planets)
in the far side-lobes of such an experiment, it is necessary to perform
the beam convolution over the full sky. We discuss this process in multipole
space for an arbitrary beam response, fully including the effects of beam
asymmetry and cross-polarization. The form of the convolution in multipole
space is such that the Wandelt-G\'{o}rski fast technique for all-sky
convolution of scalar signals (e.g.\ temperature) can be applied with little
modification. We further
show that for the special case of a pure co-polarized, axisymmetric beam
the effect of the convolution can be described by spin-weighted window
functions. In the limits of a small angle beam and large Legendre multipoles,
the spin-weight 2 window function for the linear polarization reduces to the
usual scalar window function used in previous analyses of beam effects in CMB
polarimetry experiments. While we focus on the example of polarimetry
experiments in the context of CMB studies, we emphasise that the formalism we
develop is applicable to anisotropic filtering of arbitrary tensor fields on
the sphere.
\end{abstract}

\hspace{1.5cm}
]

\section{Introduction}

Over the past decade a number of increasingly sophisticated experiments have
reported detections of the temperature anisotropy in the cosmic microwave
background (CMB). Following the detection of degree-scale anisotropy by the
COBE satellite~\cite{smoot92}, ground-based and balloon-borne experiments have
pushed back the limits on resolution and sensitivity to provide estimates of
the anisotropy power spectrum $C^{\text{T}}_l$ up to Legendre multipoles of
$l\approx 700$ (see e.g.\ Ref.~\cite{barreiro00} for a review of
the situation pre-\textsc{BOOMERANG}~\cite{debernadis00,boomerang-url} and
\textsc{MAXIMA}-1~\cite{hanay00,maxima-url}).
The anisotropy power spectrum encodes a wealth
of cosmological information (e.g.\ Ref.~\cite{hu95b}) in a highly compressed
form, making it a very convenient data product from which to determine
cosmological parameters (see Ref.~\cite{bond00} and references therein).

The combination of Thomson scattering and the non-zero temperature quadrupole,
as the radiation begins
to free stream through recombination, leads to the robust prediction that
the CMB should be linearly polarized, with an r.m.s. level of a few percent
of the temperature anisotropies~\cite{rees68,basko80,kaiser83,bond84}.
Detection of polarization would provide
complementary information to that obtained from temperature measurements,
e.g.\ the unique polarization signature of gravitational waves (and vector
modes) provides the best hope of detecting their presence at last
scattering~\cite{kamion97,seljak97}. Currently, only upper limits exist
on the degree of linear polarization
(e.g.\ Refs.~\cite{penzias65,lubin83,netterfield95}; see also
Ref.~\cite{keating98} and references therein for a recent review)
but detections should be made with the MAP~\cite{map-url} and
Planck~\cite{planck-url} satellites, several
experiments from the ground~\cite{polar-url,polatron-url,cbi-url,amiba-url},
and the flights of the MAXIPOL~\cite{maxipol-url} and enhanced
BOOMERANG~\cite{boomerang-url} balloon experiments.

In an ideal linear polarimetry experiment, a given detector is configured to
respond only to a single component of the electric field of the incident
radiation, along incident directions contained within a small solid angle
$\Delta \Omega$ (the beam size).
For such an ideal experiment, the detector measures the flux
$(I+Q)\Delta \Omega/2$ where $I$ (total intensity) and $Q$ are the Stokes
parameters of the
incident radiation along the beam direction, and the polarization basis
vectors have been chosen with the $x$ direction aligned with the polarimeter.
In practice, such ideals are never achieved, and for precision polarimetry
experiments, e.g.\ Planck, it is essential to take full account of several
beam effects on the measured signal. The polarization on the sky must be
convolved with the response pattern of the detector, which will not be
perfectly axisymmetric. Furthermore, the system will generally have some
cross-polar contamination (non-zero response to more than one polarization
component). Often the beam response pattern has
non-negligible far side-lobes which are highly polarized due to reflection and
diffraction effects in the instrument, making a full-sky convolution necessary
if the effects of stray light from bright regions (such as the sun, moon and
galaxy in CMB experiments) are to be properly accounted for. The implications
of a subset of these effects for the analysis pipeline of total power
(unpolarized) experiments have been studied recently in Ref.~\cite{wu00}.

In this paper we present a formalism that allows all such non-ideal
effects to be taken account of exactly and efficiently
in multipole space. The form of the convolution for polarized data in
multipole space is very similar to that for unpolarized radiation. This fact
allows the recently suggested algorithm of Wandelt and
G\'{o}rski~\cite{wandelt00} to be used to compute rapidly the detector
output for an arbitrary pointing direction and orientation of the detector.
The formalism presented here should prove useful in simulation and modelling
of precision polarimetry experiments, as well as in the actual analysis of
experimental data. Within the multipole formalism, it is simple to invoke
approximations, such as axisymmetry of the beam, where appropriate, to
reproduce the approximate results used in previous analyses of beam effects in
CMB polarimetry experiments~\cite{zaldarriaga98,ng99}.

The paper is arranged as follows. In Sec.~\ref{far-field} we discuss the
derivation of the detector response from its far-field radiation pattern,
and introduce
the beam response tensor. The multipole expansion of the beam in spherical
scalar and tensor harmonics is presented in Sec.~\ref{multipole}.
The scanning of the detector on the sky is described by a (time-dependent)
rotation of the
beam from some standard configuration. This rotation is most conveniently
handled in multipole space, and is described in Sec.~\ref{rotate}, where an
efficient algorithm for performing the rotation and convolution is
also described. Section~\ref{axi} deals with the case of an axisymmetric beam.
It is shown that for a certain geometry of the polarized beam response, it is
possible to describe the convolution of the beam with the incident
linear polarization in terms of spin-weight 2 window functions, 
first introduced in Ref.~\cite{ng99}.
We give the window function for Gaussian beams of arbitrary angular size, and
show that in the small-angle limit we reproduce the results of Ng and
Liu~\cite{ng99}. We summarise our discussion in
Sec.~\ref{conc}. An appendix provides some details on the
rotation properties of the tensor harmonics which are needed for
Sec.~\ref{rotate}.

\section{Far-field radiation pattern and the beam response}
\label{far-field}

It is convenient to characterise the response of the detector and feed
system by the far-field radiation pattern it would emit if used as a
transmitter, rather than a receiver~\cite{mott-antenna}. Assuming a
(quasi-)monochromatic system, the electric field in the far field is of
the form
\begin{equation}
\bbox{E} \propto \frac{1}{r} \Re\{\tilde{\bbox{\cle}}
\exp[i (kr - \omega t)]\},
\label{oldeq:1}
\end{equation}
where $\tilde{\bbox{\cle}}$ is a complex, transverse vector function on the
sphere. (We refer to geometric objects having no components outside the
surface of the sphere as being transverse.)
Here $r$ is radial distance and $\omega = c k$ is the (mean) angular frequency
of the radiation, where $k \gg 1/r$ is the wavenumber.
The most general detector and feed system will produce a partially polarized
signal in transmission, so that $\tilde{\bbox{\cle}}$ will generally be
a slowly varying function of time (compared to $\omega$). However, only
the stationary statistical properties of $\tilde{\bbox{\cle}}$ are important
for determining the response of the system to incident radiation.
With the system in some specified orientation, the power received
$\ud W_{\text{tot}}$ when illuminated by the sky along some direction
$\bbox{e}$ is proportional to the intensity in that polarization mode
which is the time reverse of Eq.~(\ref{oldeq:1}). It follows that
\begin{equation}
\ud W_{\text{tot}} \propto \langle |\bbox{\cle} \cdot \tilde{\bbox{\cle}}
|^2 \rangle \ud \Omega,
\end{equation}
where $\bbox{\cle}$ is the complex representative (or analytic signal)
of the incident electric
field propagating along $-\bbox{e}$, $\ud \Omega$ is the element of solid
angle, and angle brackets denote
time averaging over the slow variations in $\bbox{\cle}$ and
$\tilde{\bbox{\cle}}$. Writing the components of the fields $\bbox{\cle}$ and
$\tilde{\bbox{\cle}}$ on the orthonormal basis vectors
$\bbox{\sigma}_\theta$ and $\bbox{\sigma}_\phi$ of a spherical polar
coordinate system as e.g.\ $\cle_\theta$
and $\cle_\phi$, the contribution to $W_{\text{tot}}$ can be written
as
\begin{equation}
\frac{\ud W_{\text{tot}}}{\ud \Omega}
\propto \frac{1}{2} (I\tilde{I} + Q \tilde{Q} + U \tilde{U} - V \tilde{V}),
\label{eq:dwtot}
\end{equation}
where $\{I,Q,U,V\}$ are the Stokes parameters of the incoming radiation on
the $\{\bbox{\sigma}_\theta,-\bbox{\sigma}_\phi\}$ basis:
\begin{eqnarray}
I &=& \langle |\cle_\theta|^2 + |\cle_\phi|^2 \rangle, \\
Q &=& \langle |\cle_\theta|^2 - |\cle_\phi|^2 \rangle, \\
U &=& - 2 \Re \langle \cle_\theta \cle_\phi^\ast \rangle, \\
V &=& 2 \Re \langle i \cle_\theta \cle_\phi^\ast \rangle,
\end{eqnarray}
and $\{\tilde{I},\tilde{Q},\tilde{U},\tilde{V}\}$ are effective Stokes
parameters for the beam:
\begin{eqnarray}
\tilde{I} &=& \langle |\tilde{\cle}_\theta|^2 + |\tilde{\cle}_\phi|^2 \rangle
,\\
\tilde{Q} &=& \langle |\tilde{\cle}_\theta|^2 - |\tilde{\cle}_\phi|^2 \rangle
, \\
\tilde{U} &=& - 2 \Re \langle\tilde{\cle}_\theta \tilde{\cle}_\phi^\ast\rangle
, \\
\tilde{V} &=& 2 \Re \langle i\tilde{\cle}_\theta \tilde{\cle}_\phi^\ast\rangle.
\end{eqnarray}
Note that the effective Stokes parameters for the beam are defined on the same
basis as the incoming radiation, which is responsible for the minus sign in
front of the $V \tilde{V}$ term in Eq.~\ref{eq:dwtot}.

The intensity $I$ and circular polarization $V$ are invariant under rotations
of the polarization basis vectors, whilst $Q$ and $U$ transform like the
components of a second-rank tensor. Introducing the linear polarization
tensor for the incident radiation
\begin{eqnarray}
\clp^{ab}(\blde) &=& \frac{1}{2}[Q(\bldsig_\theta \otimes \bldsig_\theta
- \bldsig_\phi \otimes \bldsig_\phi) \nonumber \\
&& \mbox{}\mbox{}- U (\bldsig_\theta \otimes \bldsig_\phi
+ \bldsig_\phi \otimes \bldsig_\theta)],
\end{eqnarray}
we can write the total power received in the basis-independent form
\begin{equation}
W_{\text{tot}} \propto \frac{1}{2} \int
(I\tilde{I} - V \tilde{V} + 2\clp_{ab} \clb^{ab}) \, \ud\Omega  .
\label{eq:conv1}
\end{equation}
Here, $\clb_{ab}$ is the (linear) beam response tensor:
\begin{eqnarray}
\clb^{ab}(\blde) &=&\frac{1}{2}[\tilde{Q}(\bldsig_\theta \otimes \bldsig_\theta
- \bldsig_\phi \otimes \bldsig_\phi) \nonumber \\
&& \mbox{}\mbox{}- \tilde{U} (\bldsig_\theta \otimes \bldsig_\phi +
\bldsig_\phi \otimes \bldsig_\theta)] \\
&=& \Re [\langle\tilde{\bbox{\cle}} \otimes \tilde{\bbox{\cle}}^\ast\rangle
]^{\TT}, 
\end{eqnarray}
where ${}^{\TT}$ denotes the transverse, trace-free part. Our main task
now is to derive the dependence of the total power received
on the pointing direction and orientation of the detector.

\subsection{Co- and cross-polarized basis}

The polar basis $\{\bldsig_\theta,\bldsig_\phi\}$ is fixed relative to the sky
and is singular at the north and south poles. For describing the beam, it is
standard practice to use an alternative basis which is fixed relative to
the detector, and has its only singularity in the opposite direction to the
main beam~\cite{ludwig73}. We define a set of Cartesian basis vectors
$\{ \bldsig'_x, \bldsig'_y, \bldsig'_z \}$ which are fixed relative to
detector. It is convenient to take $\bldsig'_z$ to be along the (nominal)
main beam, and $\bldsig'_y$ along the polarization direction on axis.
Using this Cartesian frame we derive a set of polar basis vectors
$\{ \bldsig'_\theta,\bldsig'_\phi \}$ on the sphere in the standard manner.
The co- and cross-polar basis vectors are then derived by parallel-transporting
$\bldsig'_y$ and $\bldsig'_x$ respectively from the north pole
along great circles through the poles:
\begin{eqnarray}
\sigco &=& \sin \phi' \,\bldsig'_\theta + \cos \phi'\, \bldsig'_\phi
\label{eq:cobasis} \\
\sigcross &=& \cos \phi'\, \bldsig'_\theta - \sin \phi'\, \bldsig'_\phi,
\end{eqnarray}
where $\theta'$ and $\phi'$ are spherical polar coordinates.
A well-defined linearly polarized
receiver has $|\etcross| \ll |\etco|$ along the main beam, where
e.g.\ $\etco$ is the component of $\tilde{\bbox{\cle}}$ along
$\bldsig_{\text{co}}$. Cross-polar contamination arises from a non-zero
$|\etcross|$. To rotate from the co- and cross-polarized basis to the spherical
polar basis we have to rotate through $\pi/2 - \phi'$ in a right-handed sense
about the inward normal to the sphere. Transforming the Stokes
parameters for the beam from the co- and cross-polar basis to the
spherical polar basis, we have:
\begin{eqnarray}
\tilde{I} &=& \langle |\etco|^2 + |\etcross|^2 \rangle, \label{eq:beam1}\\
\tilde{Q} &=& -\langle |\etco|^2 - |\etcross|^2 \rangle\cos 2\phi' \nonumber\\
&&\mbox{} + 2\Re \langle\etco\etcross^\ast\rangle \sin 2\phi' ,\label{eq:q1}\\
\tilde{U} &=& -\langle|\etco|^2 - |\etcross|^2\rangle\sin 2\phi' \nonumber\\
&&\mbox{} - 2\Re \langle \etco\etcross^\ast \rangle \cos 2\phi',\label{eq:u1}\\
\tilde{V} &=& -2 \Re \langle i \etco \etcross^\ast \rangle, \label{eq:beam4}
\end{eqnarray}
which reflect the spin-2 nature of the linear polarization.
For simulation purposes, $\tilde{\bbox{\cle}}$ is usually determined with
physical optics codes, the results being reported on the co- and cross-polar
basis. For real experiments the Stokes parameters for the beam must be
calibrated using sources with known surface brightness and polarization.
Note that we do not assume that the beam response is fully polarized, so the
formalism developed here can also be applied to total power experiments (only
$\tilde{I}$ non-zero). In practice, the optics and feeds will introduce some
beam polarization in the side-lobes even for a nominal total power experiment.
Although small, the role of such effects in total power experiments could be
quantified with our formalism.

\section{Multipole expansions}
\label{multipole}

The dependence of the total power received on the direction and orientation
of the telescope is most easily formulated in multipole space. We say that
the detector is in its reference orientation when it is oriented so that
the basis $\{\bldsig'_x , \bldsig'_y, \bldsig'_z \}$ coincides with the
$\{\bldsig_x , \bldsig_y, \bldsig_z \}$ basis, which is fixed relative to the
sky. We describe the beam via a set of constant multipole coefficients
which are extracted on the sky basis when the detector is in its reference
orientation. To describe an arbitrary orientation of the detector at some time
along the scan, we can rotate the beam (which is most easily performed in
multipole space) to obtain the rotated beam response which is convolved with
the sky, as in Eq.~(\ref{eq:conv1}).

In the reference orientation, the total intensity and circular polarization
parts of the beam response can
be expanded in scalar spherical harmonics, e.g.\
\begin{equation}
\tilde{I}(\bbox{e}) = \sum_{lm} b^{I}_{(lm)} Y_{(lm)}(\bbox{e}),
\end{equation}
where the sum is over $l\geq 0$ and $|m| \leq l$. The multipoles $b^V_{(lm)}$
for the circular polarization are defined analogously. For the beam response
tensor $\clb_{ab}$ we must expand in the transverse, trace-free tensor
harmonics. Here we follow the coordinate-dependent approach of
Ref.~\cite{kamion97} (although for some applications the coordinate-free
approach of Ref.~\cite{chall99c} is more convenient):
\begin{equation}
\clb_{ab}(\blde) = \sum_{\text{P}lm} \bp_{(lm)} \yp_{(lm)ab},
\end{equation}
where the sum is over $l\geq 2$, $|m|\leq l$, and the two types of
transverse trace-free harmonics $\text{P}=\text{G}$ (for Gradient, often
called electric) or $\text{C}$ (Curl, often called magnetic). All
multipoles satisfy $b_{(lm)}^\ast = (-1)^m b_{(l-m)}$ since the fields are
real and we have adopted the Condon-Shortley phase for the spherical
harmonics. The Stokes parameters $I$ and $V$ for the sky can be similarly
expanded in multipoles $a^I_{(lm)}$ and $a^V_{(lm)}$, and the linear
polarization in multipoles $\ag_{(lm)}$ and $\ac_{(lm)}$.

The tensor harmonics are derived from the scalar harmonics by covariant
differentiation over the sphere~\cite{kamion97} (see also the Appendix).
Performing the differentiation gives~\cite{thorne80}
\begin{eqnarray}
Y^{{\mathrm{G}}\, ab}_{(lm)} &=& \frac{1}{\sqrt{2}}
\left({}_{-2}Y_{(lm)} \bldm \otimes \bldm
+ {}_{2}Y_{(lm)} \bldm^\ast \otimes \bldm^\ast \right), \\
Y^{{\mathrm{C}}\, ab}_{(lm)} &=& \frac{1}{i\sqrt{2}}
\left({}_{-2}Y_{(lm)} \bldm \otimes \bldm
- {}_{2}Y_{(lm)} \bldm^\ast \otimes \bldm^\ast \right),
\end{eqnarray}
where $\bldm \equiv (\bldsig_\theta + i\bldsig_\phi)/\sqrt{2}$. The
${}_{\pm 2}Y_{(lm)}$ are the spin-weight 2 harmonics:
\begin{equation}
{}_{\pm 2}Y_{(lm)} = \frac{N_l}{\sqrt{2}}(W_{(lm)}\pm i X_{(lm)}),
\end{equation}
where $N_l \equiv [2(l-2)!/(l+2)!]^{1/2}$, and
\begin{eqnarray}
W_{(lm)}(\theta,\phi) &=& \frac{\partial^2}{\partial \theta^2} Y_{(lm)}
- \cot\theta \frac{\partial}{\partial \theta}Y_{(lm)} \nonumber \\
&&\mbox{} - \csc^2\theta \frac{\partial^2}{\partial \phi^2} Y_{(lm)},
\label{eq:wlm}\\
X_{(lm)}(\theta,\phi) &=& 2 \csc\theta \left(\frac{\partial^2}{\partial \theta
\partial \phi} Y_{(lm)} - \cot\theta \frac{\partial}{\partial \phi}
Y_{(lm)}\right). \label{eq:xlm}
\end{eqnarray}
Explicit expressions for $W_{(lm)}$ and $X_{(lm)}$ with the derivatives
eliminated are given in Ref.~\cite{kamion97c}. Note that our convention
for the spin-weight functions follows Ref.~\cite{ng99}, which differs from
Goldberg et al.~\cite{goldberg67} by the inclusion of the factor $(-1)^m$
in the definition of the spherical harmonics.

\subsection{Extracting the beam multipoles}

With the detector in the reference orientation, the spherical polar bases
fixed relative to the sky and detector coincide. We can extract the
beam multipoles from the effective Stokes parameters on this polar basis
using the orthonormality of the scalar and tensor harmonics
(e.g.\ Ref.~\cite{kamion97}). For the
linear polarization, we have
\begin{equation}
\bg_{(lm)} \pm i \bc_{(lm)} = \frac{1}{\sqrt{2}} \int
\, (\tilde{Q} \mp i \tilde{U}) {}_{\pm 2}Y_{(lm)}^\ast \, \ud \Omega,
\label{eq:spin1}
\end{equation}
which is the inverse of the expansion of $\tilde{Q}$ and $\tilde{U}$ in
spin-weight 2 harmonics:
\begin{equation}
\frac{1}{\sqrt{2}}(\tilde{Q} \pm i \tilde{U}) = \sum_{lm}
(\bg_{(lm)} \mp i \bc_{(lm)}) {}_{\mp 2}Y_{(lm)}.
\label{eq:spin2}
\end{equation}
(The sum is over $l\geq 2$ and $|m| \leq l$.) Note that with our conventions
for the polarization basis vectors, $\tilde{Q} \pm i \tilde{U}$ is a
spin-weight $\mp2$ quantity, which differs from some authors
(notably Ref.~\cite{seljak97}).

\begin{figure}
\begin{center}
\includegraphics[width=8cm,angle=-90]{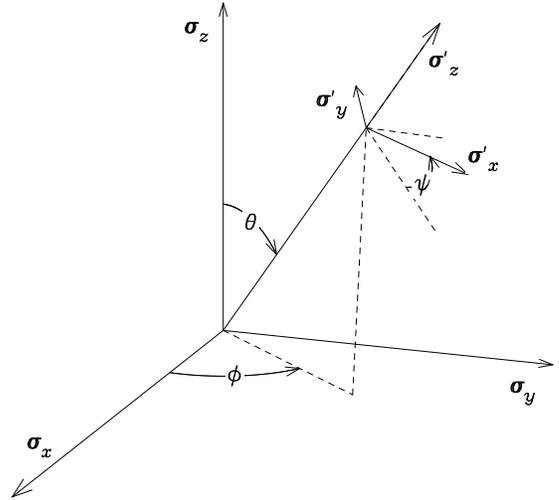}
\end{center}
\caption{The orientation of the detector is specified by the three
Euler angles $\{\phi,\theta,\psi\}$ which takes the $\{\bldsig_i\}$
frame ($i=x$, $y$, $z$), which is fixed relative to the sky, onto the
$\{\bldsig'_i\}$ frame which is fixed relative to the detector. The
nominal main beam of the detector is along $\bldsig'_z$, and the
co-polar direction along the main beam is $\bldsig'_y$.}
\label{fig:one}
\end{figure}

Some care is needed in extracting the beam multipoles at the north and
south poles since the Stokes parameters on the polar basis are ill-defined
there. However, using Eqs.~(\ref{eq:q1}) and (\ref{eq:u1}), it is simple to
show that any well-defined polarization field must have $\phi$ dependence
going like $Q \pm i U \propto \exp{(\pm 2 i \phi)}$ on the polar basis at
the north pole ($\theta=0$). This is consistent with Eq.~(\ref{eq:spin2})
since the spin-weight $\pm 2$ harmonics satisfy
\begin{equation}
{}_{\pm 2} Y_{(lm)} = \delta_{m\, \mp 2}
\sqrt{\frac{(2l+1)}{4\pi}} \text{e}^{\mp 2 i \phi}
\label{eq:spin3}
\end{equation}
at $\theta = 0$. With finitely sampled, simulated data for
$\tilde{\cle}_{\text{co}}$ and $\tilde{\cle}_{\text{cross}}$, the
contribution to $\bp_{(lm)}$ from samples on (or very near to) the north pole
can be treated by approximating ${}_{\pm2}Y_{(lm)}$ with Eq.~(\ref{eq:spin3}),
and absorbing the $\exp(\pm 2 i \phi)$ into $(\tilde{Q} \mp i \tilde{U})
\exp(\pm 2 i \phi)$ which is then well-defined by the data.
Similar problems occur at the south pole, but since the beam has virtually no
power there this is not problematic.

\section{Beam rotation and convolution}
\label{rotate}

The kinematics of the experiment can be specified by a scan strategy
which describes the rotation necessary to take each detector from the
reference orientation to its orientation at the specified time in
the scan. For simplicity we consider only a single detector, but our
approach could easily be generalised to experiments with multiple detectors.
The rotation is specified by its Euler angles $\{\phi,\theta,\psi \}$,
such that first we rotate in a right-handed sense by $\psi$ about $\bldsig_z$,
then by $\theta$ about $\bldsig_y$, and finally by $\phi$ about $\bldsig_z$
again. We denote the rotation by $D(\phi,\theta,\psi)$ so that the image of
$\bldsig_i$ is $D(\phi,\theta,\psi)\bldsig_i = \bldsig'_i$ for $i=x$, $y$,
$z$. For $\psi=0$,
\begin{eqnarray}
\bldsig'_z &=& \sin\theta (\cos\phi\, \bldsig_x + \sin\phi\, \bldsig_y)
+ \cos\theta\, \bldsig_z \nonumber \\
\bldsig'_y &=& -\sin\phi\, \bldsig_x + \cos\phi\, \bldsig_y.
\end{eqnarray}
For non-zero $\psi$, $\bldsig'_x$ and $\bldsig'_y$ are additionally
rotated by $\psi$ about $\bldsig'_z$ (see Fig.~\ref{fig:one}).

Under the rotation $D(\phi,\theta,\psi)$ the beam Stokes parameters
$\tilde{I}$ and $\tilde{V}$ rotate as scalar fields so that e.g.\
\begin{equation}
\tilde{I}(\bbox{e}) \rightarrow \tilde{I}[D(-\psi,-\theta,-\phi)\bbox{e}],
\label{eq:rot1}
\end{equation}
where $D(-\psi,-\theta,-\phi)$ is the inverse rotation. The beam response
tensor $\clb_{ab}(\bbox{e})$ rotates as a rank-two tensor, so that
\begin{equation}
\clb_{ab}(\bbox{e}) \rightarrow \Lambda_a{}^{c_1} \Lambda_b{}^{c_2}
\clb_{c_1 c_2}[D(-\psi,-\theta,-\phi)\bbox{e}],
\label{eq:rot2}
\end{equation}
where $\Lambda_a{}^{c}$ is the $\text{SO}(3)$ rotation matrix representing
$D(\phi,\theta,\psi)$ (see Appendix). Since we are describing the beams in
multipole space, we must consider the transformation properties of the
scalar and tensor harmonics under the rotations given in Eqs.~(\ref{eq:rot1})
and (\ref{eq:rot2}) respectively. The scalar spherical harmonics transform
irreducibly under rotations as~\cite{brink93}
\begin{equation}
Y_{(lm)}(\bbox{e}) \rightarrow \sum_{|m'|\leq l} D^l_{m'm}(\phi,\theta,\psi)
Y_{(lm')}(\bbox{e}),
\end{equation}
and, as we show in the Appendix, the same is true of the tensor harmonics:
\begin{equation}
Y_{(lm)ab}(\bbox{e}) \rightarrow \sum_{|m'|\leq l} D^l_{m'm}(\phi,\theta,\psi)
Y_{(lm')ab}(\bbox{e}).
\end{equation}
Here, the $D^l_{m'm}(\phi,\theta,\psi)$ are Wigner's $D$-matrices.
With our conventions for the Euler angles, we have
\begin{equation}
D^l_{m'm} (\phi,\theta,\psi) = \ue^{-im'\phi} d^l_{m'm}(\theta)
\ue^{-im\psi},
\end{equation}
where
\begin{eqnarray}
d^l_{mn}(\theta) &=& \sum_t (-1)^t \frac{[(l+m)!(l-m)!(l+n)!(l-n)!]^{1/2}}{
(l+m-t)!(l-n-t)!(t+n-m)!t!}\nonumber \\
&&\mbox{} \times [\cos(\theta/2)]^{2l+m-n-2t}
[\sin(\theta/2)]^{n-m+2t}.
\end{eqnarray}
The sum is over integers $t$ such that the arguments of the factorials are
non-negative.

Performing the integral over the sphere in Eq.~(\ref{eq:conv1})
is now straightforward in multipole space using the orthonormality of the
harmonics. The final result for the total power as a function of orientation
of the detector is
\begin{eqnarray}
W_{\text{tot}}(\phi,\theta,\psi) &\propto& \sum_{lmm'}
\biggl[\frac{1}{2} \left(a^{I \ast}_{(lm)} b^I_{(lm')}
-  a^{V \ast}_{(lm)} b^V_{(lm')}\right) \nonumber \\
&&\mbox{} + \sum_{\text{P}} a^{\text{P} \ast}_{(lm)}
\bp_{(lm')} \biggr] D^l_{mm'}(\phi,\theta,\psi).
\label{eq:result}
\end{eqnarray}
The sum is over $l \geq 0$ with $|m|$ and $|m'| \leq l$, where we defined
the linear polarization multipoles to be zero for $l=0$ and $1$. 
Our result for the dependence of the total power on orientation
is quite general; we have made no assumptions about the beam
profile and level of cross-polar contamination. Equation~(\ref{eq:result})
is one of the main results of this paper. Note that the function
$W_{\text{tot}}(\phi,\theta,\psi)$ is
expressed as a linear combination of the $D$-matrices, which form a complete
set for expanding single-valued (square-integrable) functions on the
three-sphere.

\subsection{Fast convolution algorithms}

The right-hand side of Eq.~(\ref{eq:result}) can be evaluated rapidly
by making only minor modifications to the algorithm developed recently by
Wandelt and G\'{o}rski~\cite{wandelt00} for the case of an unpolarized
detector. The key to the algorithm is to factor the rotation
$D(\phi,\theta,\psi)$ as follows:
\begin{equation}
D(\phi,\theta,\psi) = D(\phi-\pi/2,-\pi/2,\theta) D(0,\pi/2,\psi+\pi/2),
\end{equation}
so we may write
\begin{eqnarray}
D^l_{mm'}(\phi,\theta,\psi)&=&\sum_{|M|\leq l}
[D^l_{mM}(\phi-\pi/2,-\pi/2,\theta) \nonumber \\
&&\mbox{} \times D^l_{Mm'}(0,\pi/2,\psi+\pi/2)].
\end{eqnarray}
The advantage of factoring the rotation in this way is that now the
Euler angles only occur in complex exponentials, and we only need evaluate
$d^l_{mm'}(\theta)$ at $\theta=\pi/2$ [since $d^l_{mm'}(-\theta)=
d^l_{m'm}(\theta)$]. The full three-sphere of rotations can now be
calculated with a three-dimensional Fast Fourier Transform. The
$d^l_{mm'}(\theta)$ can be computed with the accurate recursive method
in Ref.~\cite{risbo96} (which can be further enhanced by making use of the
symmetries of the $d^l_{mm'}$). To perform the convolution to a resolution
corresponding to multipoles $l_{\text{max}}$, for all possible
orientations of the system, requires $O(l^4_{\text{max}})$ operations.
This should be compared with the $O(l^5_{\text{max}})$ operations required
for a brute force computation in pixel space. For experiments such as the
Planck mission, where $l_{\text{max}}$ is of the order of a few thousand, the
saving is considerable. For many experiments, the approximate azimuthal
symmetry of the beam limits the sum over $m'$ in Eq.~(\ref{eq:result}) to
$|m'| \ll l_{\text{max}}$. Since $W_{\text{tot}}$ is then a slowly varying
function of $\psi$, it is possible to sample the $\psi$ variation much more
sparsely than for $\theta$ and $\phi$, which effectively reduces the
operations count to  $O(l^3_{\text{max}})$~\cite{wandelt00}.

\section{Axisymmetric beams}
\label{axi}

In this section we consider the limiting form of the general result,
Eq.~(\ref{eq:result}), when the beam is approximated as being pure
co-polarized (i.e.\ having no cross-polar contamination), and
axisymmetric. Typically, these approximations hold well across the main
beam for a well-defined linear polarimetry system. However, the approximations
are not valid for describing the response of the system in the far
side-lobes, where the complete expression, Eq.~(\ref{eq:result}), should
be used. To quantify the errors introduced in a given experiment by assuming
an axisymmetric beam requires detailed simulation with the apparatus
developed in the earlier sections of this paper. Techniques for propagating
these errors to the band power estimates of the power spectrum in a total
power experiment have been developed recently~\cite{wu00}, but the
extension to polarized experiments must await the detailed development
of the full analysis pipeline for polarized data.

For a pure co-polar beam, $\tilde{\cle}_{\text{cross}}$ vanishes in
Eqs.~(\ref{eq:beam1})--(\ref{eq:beam4}). If we further assume axisymmetry,
then $\tilde{\cle}_{\text{co}}$ is a function of $\theta'$ alone.
Writing $|\tilde{\cle}_{\text{co}}|^2=B(\theta')$, we have
$\tilde{I} = B(\theta')$, $\tilde{V} = 0$, and
\begin{equation}
\tilde{Q} \pm i \tilde{U} = - B(\theta') \ue^{\pm 2 i \phi'}.
\end{equation}
It follows that $\bp_{(lm)}=0$ unless $|m|=2$. Furthermore, using
Eq.~(\ref{eq:spin1}) with $m=2$, we see that $\bc_{(l2)}=i\bg_{(l2)}$ and
hence $\bc_{(l\, -2)}=- i\bg_{(l\, -2)}$. The beam response tensor for linear
polarization, $\clb_{ab}$, is now fully specified by $\bg_{(l2)}$. If we use
Eqs.~(\ref{eq:wlm}) and (\ref{eq:xlm}) to express the spin-weight harmonics
in terms of Legendre functions, we find the following expression for
$\bg_{(l2)}$:
\begin{eqnarray}
\bg_{(l2)} &=& - \frac{\pi N_l^2}{\sqrt{2}} \sqrt{\frac{2l+1}{4\pi}}
\int_{-1}^1 \ud x \, B(x)\biggl\{ (l+2)(x-2)P_{l-1}^{\prime\prime} \nonumber \\
&&\mbox{}+\left[2(l-1)x - \frac{1}{2}l(l-1)(1-x^2)-(l-4)\right]
P_l^{\prime\prime} \biggr\},
\nonumber \\
&&
\end{eqnarray}
where primes denote differentiation with respect to $x\equiv \cos\theta'$,
and $P_l(x)$ is the $l$th Legendre polynomial. Note that $\bg_{(l2)}$ is real,
so $\bc_{(l2)}$ is imaginary. For the intensity multipoles we have
$b^I_{(lm)} = 0$ unless $m=0$, with
\begin{equation}
b^I_{(l0)} = 2\pi \sqrt{\frac{2l+1}{4\pi}} \int_{-1}^1 P_l(x) B(x) \, \ud x.
\end{equation}

Given the restrictions on the beam multipoles for an axisymmetric, co-polar
beam, it is possible to simplify Eq.~(\ref{eq:result}) for the total power
received. Using the relation~\cite{goldberg67}
\begin{equation}
D^l_{m\, -s}(\phi,\theta,\psi) = (-1)^s \sqrt{\frac{4\pi}{2l+1}}
{}_s Y_{(lm)}^\ast (\theta,\phi) \ue^{i s \psi}
\end{equation}
between the $D$-matrices and the spin-weight $s$ harmonics for integer
$l$, $m$, and $s$, and the reality of $b^I_{(l0)}$ and $\bg_{(l2)}$,
Eq.~(\ref{eq:result}) can be written in the form
\begin{equation}
W_{\text{tot}}(\theta,\phi,\psi) \propto \frac{1}{2} [I_{\text{eff}}
- Q_{\text{eff}} \cos 2\psi\, + U_{\text{eff}} \sin 2\psi\,]. 
\label{eq:approx}
\end{equation}
Here, $I_{\text{eff}}$ is the usual beam smoothed intensity:
\begin{equation}
I_{\text{eff}}(\theta,\phi) = \sum_{lm} W_l a^I_{(lm)} Y_{(lm)}(\theta,\phi), 
\end{equation}
where the window function
\begin{equation}
W_l = \sqrt{\frac{4\pi}{2l+1}} b^I_{(l0)}.
\end{equation}
Similarly, $Q_{\text{eff}}$ and $U_{\text{eff}}$ are beam smoothed
Stokes parameters on the spherical-polar basis, given by
\begin{equation}
\frac{1}{\sqrt{2}}(Q_{\text{eff}} \pm i U_{\text{eff}}) = \sum_{lm} {}_2 W_l
(\ag_{(lm)} \mp i \ac_{(lm)}) {}_{\mp 2}Y_{(lm)},
\end{equation}
where, following Ref.~\cite{ng99}, we have introduced the spin-weight
2 window function
\begin{equation}
{}_2 W_l = - 2 \sqrt{2} \sqrt{\frac{4\pi}{2l+1}} \bg_{(l2)}.
\label{eq:spinwindow}
\end{equation}
Note that Eq.~(\ref{eq:approx}) shows that the signal obtained by convolving
the pure co-polar, axisymmetric beam with the sky is equivalent to
the response of an idealised co-polar detector, with vanishing beam width,
on the smoothed sky. This result does not depend on any assumptions about the
angular size of the beam response; for the polarized contribution it is a
consequence of the definition of the co-polar vector field,
Eq.~(\ref{eq:cobasis}), which is the obvious generalisation of a constant
vector field to the surface of sphere.
Equations~(\ref{eq:approx})--(\ref{eq:spinwindow}) provide a complete
description of the power received in polarimetry experiments with
axisymmetric, co-polar beams.

\subsection{Gaussian beams}

It is often the case that the axisymmetric beam profile is approximately
Gaussian:
\begin{equation}
B(\theta') = B \exp[-(1-\cos\theta'\, )/\sigma^2],
\label{eq:beamprofile}
\end{equation}
where $\sigma$ is a measure of the beam width. Note that we follow
Bond and Efstathiou~\cite{bond87} in taking the beam to be Gaussian in
$2\sin(\theta'/2)$ rather than $\theta'$. The former allows us to derive simple
analytic results valid for any $\sigma$. However, in most cases a Gaussian
profile is only appropriate close to the beam axis, in which case the
two definitions are almost indistinguishable.
For the beam profile in Eq.~(\ref{eq:beamprofile}), the
window functions $W_l$ and ${}_2 W_l$ can be evaluated analytically:
\begin{equation}
W_l=4\pi B \ue^{-\alpha} i_l(\alpha), 
\end{equation}
and
\begin{eqnarray}
{}_2 W_l &=& 2\pi B N_l^2\ue^{-\alpha}\{ 2(-1)^l[(l+2)(l-1)+6 \alpha]
\ue^{-\alpha} \nonumber \\
&&\mbox{} + [(l^2 - 4 \alpha)(l-1)^2 + 12 \alpha^2]
i_l(\alpha) \nonumber \\
&&\mbox{} + 4 \alpha(l^2 + l + 1 - 3\alpha) i_{l-1} (\alpha) \},
\end{eqnarray}
where $i_l(x)$ is a modified spherical Bessel function, and
$\alpha \equiv 1/\sigma^2$. For
$\sigma^{2} \ll 1$, which is always the case for high resolution
experiments, the window functions are very well approximated
by their asymptotic expansions, which give
\begin{eqnarray}
W_l &\approx & 2\pi B \sigma^2 \exp[-l(l+1) \sigma^2/2], \\
{}_2 W_l &\approx & 2\pi B \sigma^2 \exp\{-[l(l+1)-4] \sigma^2/2\},
\end{eqnarray}
in full agreement with Ng and Liu~\cite{ng99}. For $l\gg 1$, the
polarized and unpolarized window functions in the small-scale limit
are approximately equal, which can easily be verified by making a flat sky
approximation.

\section{Conclusion}
\label{conc}

In this paper we have presented a multipole method for describing the
response of an arbitrary detector and feed system. We fully include the
effects of non-zero beam size, asymmetric beam patterns, and cross-polar
contamination. Such inclusions are essential for the accurate modelling
and interpretation of precision polarimetry data, such as that expected
from the Planck mission.
Working in multipole space, we derived a simple expression
for the response of the system, when convolved with the sky,
as a function of the three Euler angles needed to describe a general
orientation of the system. Given the form of this expression, it is
straightforward to modify the fast algorithm of Wandelt and
G\'{o}rski~\cite{wandelt00}
to compute the system response for the entire three-sphere of
orientations in $O(l_{\text{max}}^4)$ operations. Finally, we showed
how, for the case of a pure co-polar, axisymmetric beam, the response can be
described by spin-weighted window functions. This extended the results of
Ref.~\cite{ng99} to arbitrary size beams, and we gave the exact form of the
window functions for a beam with a Gaussian profile. Although our discussion
has been in the context of CMB polarimetry experiments, the formalism
introduced here should be useful in any applications involving anisotropic
filtering of tensor fields on the sphere.

The techniques described in this paper have now been implemented in the
simulation pipeline for polarized channels of the Planck mission.

\section*{Acknowledgments}

AC acknowledges a Research Fellowship from Queens' College, Cambridge.
PF is supported by a Research Fellowship from ESA.
DM and MA are supported by PPARC. We thank the members of the Cambridge Planck
Analysis Centre (CPAC) for many useful discussions.

\appendix
\section{Rotating the tensor harmonics}

In this appendix we establish the transformation properties of the
transverse, trace-free tensor harmonics under active rotations. The
action of the rotation $D(\phi,\theta,\psi)$ can be represented
by a rotation matrix $\Lambda^a{}_b$ such that an arbitrary vector $v^a$
rotates to $v^{\prime a} = \Lambda^a{}_b v^b$. Orthogonality of the rotation
implies $\Lambda^a{}_b \Lambda_a{}^c = \delta_b^c$, so the inverse rotation
is $v^a = v^{\prime b}\Lambda_b{}^a$. Under an active rotation of the beam,
the beam response tensor rotates to $\clb'_{ab}(e^c)$ which is obtained by
forward rotating the original tensor evaluated at the back rotated position:
\begin{equation}
\clb'_{ab}(e^c) = \Lambda_a{}^{c_1} \Lambda_b{}^{c_2} \clb_{c_1 c_2}(e^d
\Lambda_d{}^c).
\end{equation}
Under this transformation, the transverse nature of the tensor field
$\clb_{ab}(e^c)$ is preserved. We shall demand the same transformation
properties for the tensor harmonics $\yp_{(lm)ab}$.

It is convenient to write the tensor harmonics in terms of covariant
derivatives on the sphere of the scalar harmonics (e.g.\ Ref.~\cite{kamion97}):
\begin{eqnarray}
\yg_{(lm)ab} &=& N_l \left(\tilde{\nabla}_a \tilde{\nabla}_b Y_{(lm)}
- \frac{1}{2} \tilde{g}_{ab} \tilde{\nabla}^2 Y_{(lm)} \right) \label{eq:ap1}\\
\yc_{(lm)ab} &=& \frac{N_l}{2}(\tilde{\epsilon}^c{}_b \tilde{\nabla}_a
\tilde{\nabla}_c Y_{(lm)} + \tilde{\epsilon}^c{}_a \tilde{\nabla}_b
\tilde{\nabla}_c Y_{(lm)}), \label{eq:ap2}
\end{eqnarray}
where $\tilde{\nabla}_a$ is the covariant derivative on the unit sphere,
$\tilde{g}_{ab}=g_{ab} - e_a e_b$ is the (induced) metric (with
$g_{ab}$ the metric of Euclidean 3-space) and
$\tilde{\epsilon}_{ab}=\epsilon_{abc}e^c$ is the projected alternating tensor.
The covariant derivative $\tilde{\nabla}_a$ is obtained from the 3-dimensional
(flat) covariant derivative $\nabla_a$ by total projection:
\begin{equation}
\tilde{\nabla}_a T_{b \dots c} = \tilde{g}^{d}_a \tilde{g}^e_b \dots 
\tilde{g}^f_c \nabla_d T_{e \dots f},
\end{equation}
for an arbitrary tensor $T_{b \dots c}$. Making use of the results
\begin{eqnarray}
(\tilde{\nabla}_a Y'_{(lm)})|_{e^c} &=& \Lambda_a{}^b (\tilde{\nabla}_b
Y_{(lm)})|_{e^d \Lambda_d{}^c} \\
(\tilde{\nabla}_a \tilde{\nabla}_b Y'_{(lm)})|_{e^c} &=&
\Lambda_a{}^{d_1} \Lambda_b{}^{d_2} (\tilde{\nabla}_{d_1}
\tilde{\nabla}_{d_2} Y_{(lm)})|_{e^{d_3} \Lambda_{d_3}{}^c},
\end{eqnarray}
it is straightforward to prove that the rotated tensor harmonics
are obtained by replacing the scalar harmonics by their rotated
counterparts $Y'_{(lm)}(e^c)= Y_{(lm)}(e^d \Lambda_d{}^c)$ in
Eqs.~(\ref{eq:ap1}) and (\ref{eq:ap2}). Since the $l$th scalar harmonics
transform irreducibly under rotations as (e.g.\ Ref.~\cite{brink93})
\begin{equation}
Y'_{(lm)}(\bbox{e}) = \sum_{|m'|\leq l} D^l_{m'm}(\phi,\theta,\psi)
Y_{(lm')}(\bbox{e}),
\end{equation}
the tensor harmonics inherit the same transformation law:
\begin{equation}
Y^{\text{P}'}_{(lm)}(\bbox{e}) = \sum_{|m'|\leq l} D^l_{m'm}(\phi,\theta,
\psi) \yp_{(lm')}(\bbox{e}).
\end{equation}


\begin{thebibliography}{10}

\bibitem{smoot92}
G.~F.~Smoot \emph{et al.}, Astrophys. J. {\bf 396},  {L}1  (1992).

\bibitem{barreiro00}
R.~B. Barreiro, New. Astron. Rev. {\bf 44}, 179 (2000).

\bibitem{debernadis00}
P. de~Bernandis~et al., Nature {\bf 404},  955  (2000).

\bibitem{boomerang-url}
\begin{verbatim}http://www.physics.ucsb.edu/~boomerang/.\end{verbatim}

\bibitem{hanay00}
S.~Hanay et~al., astro-ph/0005123 (unpublished).

\bibitem{maxima-url}
\begin{verbatim}http://www.cfpa.berkeley.edu/group/cmb/.\end{verbatim}

\bibitem{hu95b}
W. Hu and N. Sugiyama, Astrophys. J. {\bf 444},  489  (1995).

\bibitem{bond00}
J.~R. Bond, A.~H. Jaffe, and L. Knox, Astrophys. J {\bf 533},  19  (2000).

\bibitem{rees68}
M.~J. Rees, Astrophys. J. Lett. {\bf 153},  1  (1968).

\bibitem{basko80}
M.~M. Basko and A.~G. Polnarev, Sov. Astron. {\bf 24},  268  (1980).

\bibitem{kaiser83}
N. Kaiser, Mon. Not. R. Astron. Soc. {\bf 202},  1169  (1983).

\bibitem{bond84}
J.~R. Bond and G. Efstathiou, Astrophys. J. Lett. {\bf 285},  45  (1984).

\bibitem{kamion97}
M. Kamionkowski, A. Kosowsky, and A. Stebbins, Phys. Rev. Lett. {\bf 78},  2058
   (1997).

\bibitem{seljak97}
U. Seljak and M. Zaldarriaga, Phys. Rev. Lett. {\bf 78},  2054  (1997).

\bibitem{penzias65}
A.~A. Penzias and R.~W. Wilson, Astrophys. J. {\bf 142},  419  (1965).

\bibitem{lubin83}
P. Lubin, P. Melese, and G. Smoot, Astrophys. J. Lett. {\bf 273},  51  (1983).

\bibitem{netterfield95}
C.~B. Netterfield {\it et~al.}, Astrophys. J. Lett. {\bf 445},  69  (1995).

\bibitem{keating98}
B. Keating, P. Timbie, A. Polnarev, and J. Steinberger, Astrophys. J. {\bf
  495},  580  (1998).

\bibitem{map-url}
\begin{verbatim}http://map.gsfc.nasa.gov/.\end{verbatim}

\bibitem{planck-url}
\begin{verbatim}http://astro.estec.esa.nl/Planck/.\end{verbatim}

\bibitem{polar-url}
\begin{verbatim}http://cmb.physics.wisc.edu/polar/.\end{verbatim}

\bibitem{polatron-url}
\begin{verbatim}http://astro.caltech.edu/~lgg/polatron/ppro.html.\end{verbatim}

\bibitem{cbi-url}
\begin{verbatim}http://astro.caltech.edu/~tjp/CBI/.\end{verbatim}

\bibitem{amiba-url}
\begin{verbatim}http://www.asiaa.sinica.edu.tw/amiba/.\end{verbatim}

\bibitem{maxipol-url}
\begin{verbatim}http://www.physics.umn.edu/cosmology/maxipol/.\end{verbatim}

\bibitem{wu00}
J.~H.~P. Wu {\it et~al.}, astro-ph/0007212 (unpublished)

\bibitem{wandelt00}
B.~D. Wandelt and K.~M. G\'{o}rski, astro-ph/0008227 (unpublished) 

\bibitem{zaldarriaga98}
M. Zaldarriaga, Astrophys. J. {\bf 503},  1  (1998).

\bibitem{ng99}
K. Ng and G. Liu, Int. J. Mod. Phys. D {\bf 8},  61  (1999).

\bibitem{mott-antenna}
H. Mott, {\em Antennas for radar and communications} (Wiley, New York, 1992).

\bibitem{ludwig73}
A.~C. Ludwig, IEEE Trans. Antennas Propagat. {\bf AP-21},  116  (1973).

\bibitem{chall99c}
A.~D. Challinor, Phys. Rev. D (to be published).

\bibitem{thorne80}
K.~S. Thorne, Rev. Mod. Phys. {\bf 52},  299  (1980).

\bibitem{kamion97c}
M. Kamionkowski, A. Kosowsky, and A. Stebbins, Phys. Rev. D {\bf 55},  7368
  (1997).

\bibitem{goldberg67}
J.~N. Goldberg {\it et~al.}, J. Math. Phys. {\bf 8},  2155  (1967).

\bibitem{brink93}
D.~M. Brink and G.~R. Satchler, {\em Angular Momentum}, 3rd  ed. (Clarendon
  Press, Oxford, 1993).

\bibitem{risbo96}
T. Risbo, J. Geodesy {\bf 70},  383  (1996).

\bibitem{bond87}
J.~R. Bond and G. Efstathiou, Mon. Not. R. Astron. Soc. {\bf 226},  655
  (1987).

\end{thebibliography}
\end{document}